\documentstyle[preprint,aps]{revtex}

\newcommand{\unit}{\hat{\bf n}}

\newcommand{\rv}{{\bf r}}
\newcommand{\Ev}{{\bf E}}

\newcommand{\dv}{{\bf d}}

\newcommand{\kv}{{\bf k}}

\newcommand{\beq}{\begin{equation}}
\newcommand{\eeq}{\end{equation}}
\newcommand{\bea}{\begin{eqnarray}}
\newcommand{\eea}{\end{eqnarray}}

\begin{document}

\draft
\preprint{}
\title{Macroscopic superpositions of Bose-Einstein
condensates}
\author{Janne Ruostekoski$^1$, M. J. Collett$^1$, Robert Graham$^2$, and Dan F.
Walls$^1$}
\address{ $^1$Department of Physics, University of Auckland, Private Bag
92019, \\Auckland, New Zealand\\
$^2$Universit\"at GH Essen, Fachbereich Physik, D45117 Essen, Germany}
\date{\today}
\maketitle
\begin{abstract}
We consider two dilute gas Bose-Einstein condensates with opposite
velocities from which a monochromatic light field detuned far from the
resonance of the optical transition is coherently scattered. In the
thermodynamic
limit, when the relative fluctuations of the atom number difference between the
two condensates vanish, the relative phase between the Bose-Einstein condensates
may be established in a superposition state by detections of spontaneously
scattered photons, even though the condensates have initially well-defined atom
numbers. For a finite system, stochastic simulations show that
the measurements of the scattered photons lead to a randomly drifting
relative phase and drive the condensates into entangled superpositions of number
states. This is because according to Bose-Einstein statistics the scattering
to an already occupied state is enhanced.
\end{abstract}

\section{Introduction}

Bose-Einstein condensates (BECs) of ultra-cold trapped atomic gases
\cite{AND95,DAV95,MEW96,BRA97} have stimulated interest on the coherence
properties of matter. BECs are expected to exhibit a macroscopic quantum
coherence which in thermal atomic ensembles is absent. Even BECs with a
well-defined number of atoms, and with no phase information, could show phase
correlations in particular measurement processes on atoms
\cite{JAV96,NAR96,CIR96,JAC96,WON96,CAS97,STE97}, or on photons
\cite{RUO97d}. The relative phase between two BECs could be determined, for
instance, by various optical methods \cite{RUO97d,JAV96b,IMA96,RUO97b,SAV97}. In
recent experiments Andrews {\it et al.} \cite{AND97} have found  evidence of
macroscopic quantum coherence in a BEC by measurements of the interference
of two
condensates by absorption imaging. The two independent and spatially separated
BECs were created by a repulsive optical force in the center of the trap.

In this paper we consider an optical analogue of the atom detection schemes
of two BECs in different momentum states \cite{JAV96,NAR96,CIR96,WON96}. Instead
of looking at the spatial interference pattern we combine the scattered photons
from the atomic transitions between different BECs with a photon beamsplitter.
The measurements on scattered light has evident advantages over atom counting
from a theoretical point of view. In the case of light scattering we can use the
well-known theories of photon detection \cite{GLA63}. Also, the measurement of
spontaneously scattered photons is nondestructive for the condensates, because
only light is scattered and no atoms are removed from the two BECs. In our
measurement scheme it is shown via the simulations of stochastic Schr\"odinger
equations that the detections of spontaneously scattered photons drive the
condensates into macroscopic quantum superpositions of phase and number states
(``Schr\"odinger cats"). The phase superpositions are a consequence of the
particular measurement process, which is insensitive to certain phase values.
The entangled number state superpositions follow from the properties of
Bose-Einstein statistics and from the macroscopic quantum coherence of BECs. The
number state superpositions are multiparticle quantum states with spatially
nonlocal correlations. Due to the large  fluctuations of the number difference
between the BECs the relative phase  drifts randomly in the case of a finite
system and no stable phase is built up by measurements. The fluctuations of the
number difference vanish in the thermodynamic limit and the measurements
establish a stable phase.

Recently, Cirac {\it et al.} \cite{CIR97} have studied the ground state of two
coupled BECs by variational techniques. They have found that under certain
conditions the state with a minimum energy corresponds to a macroscopic
superposition of number states.

We begin in Sec.~{\ref{basic}} by introducing basic relations. In the limit
of large detuning of the driving light field from the atomic resonance the
excited state operators may be eliminated adiabatically. We obtain an effective
two-state Hamiltonian coupling the two BECs. We study the dynamics of the
system
in terms of stochastic trajectories of state vectors. In Sec.~{\ref{thermo}} we
consider the thermodynamic limit, where the fluctuations of the number
difference between the BECs vanish. The results of simulations for a finite
system are presented in Sec.~{\ref{nume}}. In Sec.~{\ref{detec}} we show that
the number state superpositions could be detected by considering the intensity
correlations of the scattered light. Finally, a few concluding remarks are made
in Sec.~{\ref{conc}}.

\section{Open system dynamics}

\subsection{Basic relations}
\label{basic}

The internal quantum state for both condensates is denoted by
$|g\rangle$. This state is optically coupled to the electronically excited state
$|e\rangle$ by the driving electric field $\Ev_d$ with a dominant frequency
$\Omega$. The light field is assumed to be in a coherent state and detuned
far from the resonance of the atomic transition. The two BECs are assumed to be
optically thin \cite{JAV95b} and
in the momentum states $\kv_0$ and $-\kv_0$. We consider the situation in which
the condensates are overlapping when the light is switched on. We only
consider the coherent spontaneous scattering between the condensates, which is
stimulated by a large number of atoms in the condensates. By spontaneous
scattering we mean that the emission is not stimulated by light, although it is
stimulated by atoms.
The decay into non-condensate center-of-mass (c.m.) states is also stimulated by
the Bose-Einstein statistics. However, at very low temperatures this stimulation
is much weaker because most of the particles are in the condensates. In addition
to the Bose stimulation of spontaneous emission there is unstimulated free-space
decay, at rate $\gamma$, which is always present. With a sufficiently large
number of atoms in the two BECs the free-space decay may be ignored.

The annihilation operators for the two BECs are
$g_{\kv_0}$ and $g_{-\kv_0}$. Here $g_{\kv_0}$ denotes the annihilation
operator for the electronic ground state $|g\rangle$ and the
c.m. state $\kv_0$ with the corresponding wave function $\phi_{g,\kv_0}(\rv)$.
To simplify the notation we define $b\equiv g_{\kv_0}$, $\phi_b(\rv) \equiv
\phi_{g,\kv_0}(\rv)$, $c\equiv g_{-\kv_0}$, and $\phi_c(\rv) \equiv
\phi_{g,-\kv_0}(\rv)$. We obtain for the Hamiltonian \cite{JAV95b,RUO97a}
\bea
H &=& \hbar\epsilon^g_{\kv_0}\,b^\dagger b+\hbar\epsilon^g_{-\kv_0}\,c^\dagger
c+ \sum_{\kv}\hbar(\omega_{eg}+\epsilon^e_{\kv})\, e_{\kv}^\dagger e_{\kv}
+\sum_q\hbar\omega_q\, a^\dagger_q a_q\nonumber\\ &-&\sum_{\kv}\left(\int
d^3r\,\dv_{ge}\cdot\Ev(\rv) \phi^*_b(\rv)\phi_{e\kv}(\rv) \,b^\dagger
e_{\kv}+ {\rm H.c.}\right)\nonumber\\ &-&\sum_{\kv}\left(\int
d^3r\,\dv_{ge}\cdot\Ev(\rv) \phi^*_c(\rv)\phi_{e\kv}(\rv)\, c^\dagger
e_{\kv}+{\rm H.c.}\right)\,, \label{eq:HDN}
\eea
where the excited state wave function for the c.m. state $\kv$ is $\phi_{e\kv}$.
The dispersion relations for the ground state and excited state c.m. frequencies
are $\epsilon^g_{\kv}$ and $\epsilon^e_{\kv}$, respectively. The photon
annihilation
operator for the mode $q$ is $a_q$. The internal atomic energy is described by
the frequency $\omega_{eg}$ of the optical transition  between the electronic
ground state and excited state. The last two terms in Eq.~{(\ref{eq:HDN})} are
for the atom-light dipole interaction. The dipole matrix element for the atomic
transition $e\rightarrow g$ is given by $\dv_{ge}$. We consider the
translationally invariant system, where the eigenfunctions for the condensates
are plane waves: $\phi_b(\rv)=e^{i\kv_0\cdot\rv}/\sqrt{V}$ and
$\phi_c(\rv)=e^{-i\kv_0\cdot\rv}/\sqrt{V}$. The driving electric field is also
described by a plane wave ${\bf E}^+_{d}({\bf r}) ={\cal E} \hat{\bf e}\,
e^{i(\kv\cdot\rv-\Omega t)}/2$.

In the limit of large detuning, $\Delta=\Omega-\omega_{eg}$, the excited state
operators $e_{\kv}$ in Eq.~{(\ref{eq:HDN})} may be eliminated adiabatically, and
the c.m. energies of the excited state may be ignored \cite{JAV95b}. The system
may then be described by an effective two-state Hamiltonian:
\bea
H &=& \hbar\epsilon^g_{\kv_0}\,b^\dagger b+\hbar\epsilon^g_{-\kv_0}\,c^\dagger
c+\sum_q\hbar\omega_q\, a^\dagger_q a_q\nonumber\\
&-&{1\over\hbar\Delta}\left\{\hat{N}
\int d^3r\,\dv_{ge}\cdot{\bf E}(\rv)\,\dv_{eg}\cdot{\bf E}
(\rv)\phi^*_b(\rv)\phi_b(\rv)\right.\nonumber\\ &&+\left.\left(b^\dagger
c\int d^3r\,\dv_{ge}\cdot{\bf E}(\rv)\,\dv_{eg}\cdot{\bf
E}(\rv)\phi^*_b(\rv) \phi_c(\rv)+{\rm H.c.}\right)\right\}\,.
\label{eq:HDN2}
\eea
Here we have used the fact that for the plane waves
$\phi^*_b(\rv)\phi_b(\rv)=\phi^*_c(\rv) \phi_c(\rv)$. The total atom number
operator
is given by $\hat{N}= b^\dagger b+ {c}^\dagger{c}$. Because the total number is
conserved, the operator $\hat{N}$ contributes to the  measurements only
through a
constant phase shift. Thus, we may ignore the term proportional to $\hat{N}$ in
Eq.~{(\ref{eq:HDN2})}.

We consider the dynamics of the two BECs and the driving light field as an
open quantum system and eliminate the vacuum electromagnetic fields.
The set-up of our {\it Gedanken} experiment is given in Fig.~{\ref{fig:1}}. The
incoming light field is scattered from two overlapping BECs moving with opposite
velocities. The scattering processes in which an atom scatters back to the same
condensate introduce the term proportional to the total number of atoms in
Eq.~{(\ref{eq:HDN2})}, and they may be ignored. In the scattering processes in
which atoms scatter between different condensates the light beams are deflected
due to the recoil momentum. In Fig.~{\ref{fig:1}} a photon deflected to left
corresponds to the change of the momentum of an atom from $-\kv_0$ to $\kv_0$,
{\it i.e.} the amplitude of the scattered electric field is proportional to
$b^\dagger c$. Similarly, a photon deflected to right corresponds to the change
of the  momentum of an atom upon scattering from $\kv_0$ to $-\kv_0$. In
this case
the amplitude of the scattered electric field is proportional to $c^\dagger b$.
The scattered light beams are combined by perfectly reflecting mirrors and a
50-50 photon beamsplitter. The detection rate of photons on the detectors is
the intensity of the scattered light $I(r)=2c\epsilon_0\langle\Ev^-(\rv)
\cdot\Ev^+(\rv)\rangle$ integrated over the scattering directions divided by the
energy of a photon $\hbar c k$. Writing the electric fields in the far radiation
zone ($kr\gg1$) \cite{JAV95b} we obtain the detection rate at the channel $j$:
\bea
\gamma_j &=& {1\over \hbar c k}\int d\Omega_{\unit}\, r^2
I_j(\rv) =2 \Gamma\langle C_j^\dagger C_j \rangle\,, \\
\Gamma &\equiv& {3\gamma\over 16\pi\hbar^2\Delta^2}\,|d_{eg}{\cal E}|^2\,,
\eea
where the linewidth of the electric dipole transition is given by
$\gamma = d_{eg}^2 k^3/ (6\pi\hbar\epsilon_0)$. The two relaxation channels
corresponding to the two output channels of the beamsplitter are
\beq
C_1 =  \hat{J}_x,\quad
C_2 =  \hat{J}_y\,,
\label{eq:relaxch}
\eeq
where the familiar angular momentum operators obeying SU(2) algebra are defined
by
\begin{mathletters}
\bea
\hat{J}_x &=&  {1\over 2} \left(  b^\dagger {c}+c^\dagger b\right)\,,\\
\hat{J}_y &=& {1\over 2i} \left(  b^\dagger {c}-c^\dagger b\right)\,,\\
\hat{J}_z &=& {1\over 2} \left(  b^\dagger {b}-c^\dagger c\right)\,,
\eea
\end{mathletters}
and $\hat{J}^2=\hat{J}_x^2+\hat{J}_y^2+\hat{J}_z^2=(\hat{N}/2+1)\hat{N}/2$ is
the Casimir invariant.

The system Hamiltonian for the BECs and for the driving electromagnetic field
is eliminated completely in the interaction representation and according to
Ref.~{\cite{RUO97d}} we may then write down
the equation of motion for the reduced density matrix of the system in the limit
of large detuning of the driving light from the atomic resonance
\beq
\dot{\rho}_S =-\Gamma\sum_{i=1}^2\left(
C_i^\dagger C_i\rho_S +\rho_S C_i^\dagger C_i -2C_i \rho_S
C_i^\dagger  \right)\,.
\label{eq:mas_u}
\eeq

If we assume the condensates to be in coherent states with equal mean atom
numbers $\langle b \rangle=\sqrt{N/2}\,e^{i\varphi_b}$ and  $\langle c \rangle
=\sqrt{N/2}\,e^{i\varphi_c}$, the intensities of the scattered light in the two
channels are
\beq
I_1\propto \langle \hat{J}_x^2 \rangle \propto (\cos\varphi)^2,\quad
I_2\propto \langle \hat{J}_y^2 \rangle \propto (\sin\varphi)^2\,.
\label{eq:amb}
\eeq
Here we have defined the value of the relative phase by
$\varphi\equiv\varphi_c-\varphi_b$, where $\varphi_b$ and $\varphi_c$ are the
macroscopic phases of the condensates $b$ and $c$, respectively. There is an
ambiguity in Eq.~{(\ref{eq:amb})} between the phase values $\pm\varphi$ and
$\pi\pm\varphi$. For phase-sensitive homodyne detection this ambiguity
vanishes.

The dynamics of the density operator from Eq.~{(\ref{eq:mas_u})} may be
unraveled
into stochastic trajectories of state vectors \cite{DAL92,GAR92,CAR93}. The
procedure consists of the evolution of the system with a non-Hermitian
Hamiltonian $H_{\rm eff}$, and randomly decided quantum `jumps' corresponding to
the direct detections of spontaneously emitted photons. The system evolution is
thus conditioned on the outcome of a measurement. The non-Hermitian Hamiltonian
is obtained from  Eq.~{(\ref{eq:mas_u})}
\beq
H_{\rm eff}=-i\hbar \Gamma\sum_{j=1}^2 C^\dagger_jC_j=-i\hbar \Gamma
\left( \hat{J}^2-\hat{J}_z^2\right)\,.
\label{eq:nonher}
\eeq
The non-Hermitian Hamiltonian $H_{\rm eff}$ determines the evolution of the
state
vector $\psi_{\rm sys}(t)$. If the wave function $\psi_{\rm sys}(t)$ is
normalized, the probability that a photon from the output channel $j$ ($j=1,2$)
of the beamsplitter is detected during the time interval $[t,t+\delta t]$
is
\beq
P_j(t)=2\Gamma\langle\psi_{\rm sys}(t)\, |\, C^\dagger_jC_j\, |\, \psi_{\rm
sys}(t)\rangle\, \delta t\,.
\label{eq:prob1}
\eeq
The probability of no detections is $1-P_1-P_2$.

The implementation of the simulation algorithm is similar to
Ref.~{\cite{RUO97d}}. At the time $t_0$ we generate a quasi-random number
$\epsilon$ which is uniformly distributed between 0 and 1. We assume that the
state vector $\psi_{\rm sys}(t_0)$ at the time $t_0$ is normalized. Then we
evolve the state vector by the non-Hermitian Hamiltonian $H_{\rm eff}$
iteratively
for finite time steps $\Delta t\simeq\delta t$. At each time step $n$ we compare
$\epsilon$ to the reduced norm of the wave function, until $\langle \psi_{\rm
sys}(t_0+n\Delta t)\, |\, \psi_{\rm sys}(t_0+n\Delta t )\rangle <\epsilon$, when
the detection of a photon occurs. After the detection we generate a new
quasi-random number $\eta$. We evaluate $P_1$ and $P_2$ from
Eq.~{(\ref{eq:prob1})} at the time of the detection. If $\eta<P_1/(P_1+P_2)$ we
say the photon has been detected from channel 1. If the photon has been
observed during the time step $t\rightarrow t+\Delta t$ we take the new wave
function at $t+\Delta t$ to be
\beq
|\, \psi_{\rm
sys}(t+\Delta t)\rangle=\sqrt{2\Gamma}\,\hat{J}_x\,|\, \psi_{\rm
sys}(t)\rangle\,,
\label{eq:meas1}
\eeq
which is then normalized. Otherwise, $\eta>P_1/(P_1+P_2)$ and the photon
has been
detected from channel 2. In that case the new wave function before
the normalization reads
\beq
|\, \psi_{\rm sys}(t+\Delta t)\rangle=\sqrt{2\Gamma}\,\hat{J}_y\,|\,
\psi_{\rm sys}(t)\rangle\,. \label{eq:meas2}
\eeq
After each detection the process starts again from the beginning.

\subsection{Thermodynamic limit}
\label{thermo}

Before presenting numerical results of the simulations of the stochastic
Schr\"odinger equations, we investigate qualitatively the build-up of the
macroscopic coherence by the measurement process in the limit
$N\rightarrow\infty$. The eigenstates of the effective Hamiltonian $H_{\rm eff}$
in Eq.~{(\ref{eq:nonher})} are number states which have flat phase amplitudes.
Thus, the time evolution of $H_{\rm eff}$ does not support any particular phase
value over other values, and the relative phase between the two BECs is
determined by the distribution of the photon detections between the two output
channels of the beamsplitter. Because the two relaxation channels do not
commute,
$[\hat{J}_x,\hat{J}_y]=i\hat{J}_z$, the state of the system depends also on the
particular order in which the scattered photons are detected. This complicates
the analysis substantially. An evident simplification is to consider the
thermodynamic limit $N\rightarrow\infty$, where the relative fluctuations
of the number
difference between the BECs vanish $\langle\hat{J}_z\rangle /\langle \hat{N}
\rangle\rightarrow 0$. In the next section we consider in the numerical
simulations systems which are far away from this limit. However, it turns out
that the qualitative behaviour of the macroscopic phases is still very similar.

In the thermodynamic limit we can replace the angular
momentum operators $\hat{J}_x$ and $\hat{J}_y$ by $(\hat
N/2)\cos\hat{\varphi}$ and
$(\hat N/2)\sin\hat{\varphi}$ respectively, where $\hat{\varphi}$ is the
relative phase
operator between the two BECs. Then the two relaxation channels in
Eq.~{(\ref{eq:relaxch})} commute,
$[\cos\hat\varphi,\sin\hat\varphi]=4i\Delta\hat N/\hat N^2\rightarrow 0$,
and the relevant commutation relations are
given
by \beq
[\Delta \hat{N}, \cos{\hat{\varphi}}] = i\sin{\hat{\varphi}},\quad
[\Delta \hat{N}, \sin{\hat{\varphi}}] = -i\cos{\hat{\varphi}}\,,
\eeq
where we have written $\Delta\hat{N}\equiv\hat{J}_z$. Geometrically, the
thermodynamic limit may be understood as a restriction of the dynamics
of the angular momentum variables to the equator $\langle \hat{J}_z\rangle
\simeq0$ of the Bloch sphere, as the radius of the sphere goes to infinity.

Now we can use the procedure developed in Ref.~{\cite{CAS97}}. To simplify the
notation, we ignore the spatial dependence of the wave functions. We expand the
number state $|N/2,N/2\rangle$ in terms of the overcomplete set of phase states
\cite{LEG91}:
\beq
|\varphi\rangle_N={1\over\sqrt{2^N N!}}\,(b^\dagger e^{-i\varphi/2} +c^\dagger
e^{i\varphi/2})^N \,|0\rangle \,.
\eeq
A similar analysis to Ref.~{\cite{CAS97}} then leads to the state of the system
after $n_1$ and $n_2$ detections from output channels 1 and 2 of the
beamsplitter respectively:
\beq
|\psi(n_1,n_2)\rangle\propto \int_0^{2\pi} d\varphi\, (\cos\varphi)^{n_1}
(\sin\varphi)^{n_2}\,|\varphi \rangle_N \,.
\label{eq:casdal}
\eeq
The value of the phase $\varphi$ that maximizes the integrand satisfies the
relation $\tan^2\varphi =n_2/n_1$. If $0\leq\varphi_0\leq\pi/2$ is a solution
for the maximum amplitude, then $-\varphi_0$ and $\pi\pm\varphi_0$ are also
solutions. For $n_1,n_2\gg1$, we can express the integrand in
Eq.~{(\ref{eq:casdal})} in terms of exponential functions and expand the
exponents in a Taylor series around $\pm\varphi_0$ and $\pi\pm\varphi_0$. We
obtain \bea
|\psi(n_1,n_2)\rangle &\propto & \int_0^{2\pi} d\varphi\, \left\{ e^{-n(\varphi
-\varphi_0)^2} +(-1)^{n_1}e^{-n(\pi-\varphi-\varphi_0)^2}\right. \nonumber\\
&&\left. +(-1)^{n_1+n_2}e^{-n(\varphi+\pi-\varphi_0)^2}
+(-1)^{n_2}e^{-n(\varphi+\varphi_0)^2}\right\} \,|\varphi \rangle_N\,,
\label{eq:casdal2}
\eea
where $n=n_1+n_2$. The phase distributions are Gaussians centered at four
superposition values $\pm\varphi_0$ and $\pi\pm\varphi_0$, where $0\leq
\varphi_0\leq\pi/2$ is a solution for $\tan^2\varphi =n_2/n_1$. The phase is
well-defined with a narrow width for the Gaussian. The transition from the
binomial distribution of Eq.~{(\ref{eq:casdal})} to the normal distribution of
Eq.~{(\ref{eq:casdal2})} is just the realization of the Central Limit
Theorem for a large number of detections, while the superpositions are a
consequence of the particular detection method, which is insensitive to the
phase values $\pm\varphi$ and $\pi\pm\varphi$ according to Eq.~{(\ref{eq:amb})}.

\subsection{Numerical results}
\label{nume}

For a finite system the two relaxation channels from
Eq.~{(\ref{eq:relaxch})} do not commute and the state of the BECs depends on the
particular order in which photons from the two output channels of the
beamsplitter are detected. We have simulated the measurements of the
spontaneously scattered photons numerically for $N=200$ atoms. Even though we
start from the initial number state $N_b=N_c=100$ with no phase information, the
detections establish coherence properties for BECs similarly to
Eq.~{(\ref{eq:casdal2})}. However, the value of the phase $\varphi_0$ in
Eq.~{(\ref{eq:casdal2})} does not stabilize due to the moderate value of $N$
chosen, even for a large number of
detections. The two BECs are also in entangled number state superpositions. The
coherence properties of the BECs vary strongly even during single realizations
of the measurement process. In the extreme case the condensates approach an
entangled number state with almost all the atoms in one of the two BECs.

The emergence of the number state superpositions may be understood from the
quantum statistical properties of Bose-Einstein particles. Because the
scattering
to the non-condensate modes is ignored, the total number of atoms in the
two BECs
is conserved and the atom numbers are entangled. According to Bose-Einstein
statistics the scattering to an already occupied state is enhanced. For the
initial state $|\,N/2,N/2\,\rangle$ the probability for light to scatter atoms
between the two BECs is, in the case of spatially overlapping condensates and
in the limit of large number of atoms, approximately
proportional to $(N/2)^2$. Because the detected photons corresponding to the two
atomic transitions between the BECs are indistinguishable, the number state
distribution remains symmetric with respect to the initial state during the
scattering process. For an entangled number state $(|\,N-k,k\,\rangle+
|\,k,N-k\,\rangle )/\sqrt{2}$ the scattering probability is approximately
proportional to $(N-k)k\leq (N/2)^2$. Hence the states with unequal atom numbers
have smaller scattering rates and they are more stable. It should be pointed out
that it is not necessary to have initially equal atom numbers to obtain number
state superpositions, although the distributions are perfectly symmetric only if
the initial atom numbers are the same.

The state of the BECs may be described in terms of quasiprobability functions.
For the number state distribution of atoms $|\psi_b\rangle = \sum_n c_n
|n\rangle$ in the condensate $b$ we have evaluated the $Q$ function
\cite{WAL94}:
\beq
Q(\alpha)={ |\langle\alpha |\psi_b \rangle |^2 \over\pi} =
{e^{-|\alpha|^2}\over\pi}\left|\sum_{n=0}^N {\alpha^n c^*_n\over
\sqrt{n!}}\right|^2\,.
\label{eq:Q}
\eeq
In Figs.~{\ref{fig:2}}, \ref{fig:3}, and \ref{fig:4}, we have plotted
$|\psi_b|$, the absolute value of the wave function in the condensate $b$ in the
number state basis, and the corresponding $Q$ function at different times during
a single realization of measurements. Figures {\ref{fig:2}} and \ref{fig:3}
represent typically observed distributions, when several thousands of
detections are made. In Fig.~{\ref{fig:4}} we have a special case in which
almost all the atoms are in one of the two BECs. In Fig.~{\ref{fig:2}}a two
distinct peaks in the number distribution are clearly observed. The first peak
is centered at $N_b \simeq 30$ atoms, {\it i.e.} $N_c\simeq 170$ atoms, and the
second at $N_b\simeq 170$ atoms, {\it i.e.} $N_c\simeq 30$ atoms. Only odd
number states are occupied. This is because in each photon detection the states
with only even numbers in the atom number distribution are changed to the states
with only odd atom numbers and {\it vice versa}. In particular, for a coherent
system we may define even $|\,\alpha,+\rangle$ and odd $|\,\alpha,-\rangle$
coherent states by $|\,\alpha,\pm\rangle\propto |\,\alpha\rangle \pm
|-\alpha\rangle$ \cite{TIT66}. These are states which have only even or odd
numbers in the atom number distribution and they correspond to superposition
states with two different phase values shifted by $\pi$.

In Fig.~{\ref{fig:2}}b we have plotted the corresponding $Q$ function from
Eq.~{(\ref{eq:Q})}. The $Q$ function gives the phase-space distribution. The
amplitude and phase quadratures are denoted by $X$ and $Y$. In polar
coordinates the radius in the $xy$ plane is equal to $N_b^{1/2}$ and the polar
angle is the relative phase between the two BECs. In Fig.~{\ref{fig:2}}b it is
easy to see the two different sets of peaks corresponding to the two dominating
values in the number distribution. All the peaks are aligned parallel to the $x$
axis. This is the reason that two of the four different phase values from
Eq.~{(\ref{eq:casdal2})} are indistinguishable. Although the distribution in
Fig.~{\ref{fig:2}}a is symmetric, the number squeezing of the peak with the
larger atom number is much stronger. The fringes indicating a quantum
interference in the Wigner distributions \cite{WAL94} are absent in the $Q$
representation, so that graphs of $Q$ functions do not obviously distinguish
between pure states and statistical mixtures. However, because we are dealing
with basis vectors instead of with density matrices, it is evident that we have
a pure state.

In Fig.~{\ref{fig:3}} the distribution in the number state basis and the
corresponding $Q$ function are plotted in the same run of measurements as in
Fig.~{\ref{fig:2}}, but at different time. In Fig.~{\ref{fig:3}}a two distinct
peaks in the number distribution are not as far apart as in
Fig.~{\ref{fig:2}}a. In the $Q$ representation, in Fig.~{\ref{fig:3}}b, it is
easy to see that the value of the relative phase between the BECs is different
from Fig.~{\ref{fig:2}}. In Fig.~{\ref{fig:3}}b all the four phase values from
Eq.~{(\ref{eq:casdal2})} are clearly observed. The value of the relative phase
between the condensates wanders during the simulations and does not stabilize to
any definite value. In Fig.~{\ref{fig:4}} we have one more graph from the
same run of measurements. In this case the BECs are almost in an entangled
number state with all the atoms in one of the two condensates. Because the state
of the BECs is closer to a number state than to a coherent state, the relative
phase is not well-defined.

In the calculations we only considered the atom-stimulated scattering to the
BECs. It is not necessarily a well-justified assumption to ignore the 
unstimulated free-space decay if the condensates contain only 200 atoms. 
However, the purpose of the numerical simulations was to demonstrate the
general properties of the finite systems with a convenient computational
efficiency. In the simulations the
physical behaviour remained qualitatively the same even though the number of
atoms was significantly increased. In the real experiments BECs have typically
contained many more than 200 atoms \cite{AND95,DAV95,MEW96,BRA97}.

The interaction of the BECs with their enviroment creates dissipation and the
decoherence of the macroscopic superpositions \cite{ZUR91}. Decoherence by
amplitude damping or by phase damping has been estimated by Walls and Milburn
\cite{WAL85}. The amplitude damping corresponds to the losses of atoms from the
BECs. In this case the off-diagonal elements of the density matrix between two
coherent states may be shown to be dephased by the factor $\langle\alpha |\beta
\rangle^{1-\exp{(-\lambda t)}}$, where $\lambda$ is the loss rate for atoms. The
phase damping may, e.g., be a consequence of elastic two-body collisions.
In this case the off-diagonal elements of the density matrix between two
coherent states with unequal atom numbers $N_1$ and $N_2$ are damped by the
factor $\exp{\{ -\lambda (N_1-N_2)^2 t/2\}}$. These are examples of the
decoherence of the ensemble averages over the measurement processes. If
decoherence can be associated with measurements, evolution of single
realizations may be analyzed by stochastic evolutions of state vectors
\cite{GAR94}. Although the decoherence of number state superpositions of the
BECs may exhibit some interesting features, we do not consider this in the
present paper.

\subsection{Detection of number state superpositions}
\label{detec}

In this section we consider the detection of the number state superpositions.
The phase superpositions could in principle be measured by simply interfering
the condensates. However, the different phase superpositions correspond to
either even or odd coherent states. As explained previously, these are states
which have only even or odd numbers in the atom number distribution. Thus, in
practice the losses of atoms from the BECs could shift the fringes and wipe out
the qualitative features from the interference pattern.

The existence of the number state superpositions could be verified, for
instance, by considering the intensity correlations of the scattered light from
the two BECs. If we assume that the condensates are flying apart and that they
are already spatially separated, the amplitude of the spontaneously scattered
light field from the condensate $b$ has roughly the dependence $|\Ev^+_b|\propto
{\cal E} d_{eg}/(\hbar\Delta) \, b^\dagger b$, and from the condensate $c$
$|\Ev^+_c|\propto {\cal E} d_{eg}/(\hbar\Delta) \, c^\dagger c$ \cite{JAV95b}.
Here we have again considered only the coherent spontaneous scattering of
atoms to the BECs stimulated by large atom numbers, {\it i.e.} scattering to
the non-condensate c.m. states has been ignored. Because the BECs are now
spatially separated, only the scattering processes in which an atom scatters
back to the same condensate are included.

For a number state $|\,N/2,N/2\,\rangle$ with large $N$, the
intensities of the scattered light from the spatially separated condensates $b$
and $c$ are approximately $\langle I_b \rangle \sim \langle I_c \rangle \propto
(N/2)^2$. The intensity correlations satisfy $\langle I_b I_c \rangle \propto
(N/2)^4$.  For an entangled number state $(|\,N-k,k\,\rangle+ |\,k,N-k\,\rangle
)/\sqrt{2}$ we have $\langle I_b \rangle \sim \langle I_c \rangle \propto
\{(N-k)^2+k^2\}/2 \geq (N/2)^2$ and  $\langle I_b I_c \rangle \propto (N-k)^2
k^2\leq (N/2)^4$. An especially interesting case is the situation in which the
superpositions are far apart: $k\ll N/2$. Then, $\{(N-k)^2+k^2\}/2\simeq
2(N/2)^2$ and $  (N-k)^2 k^2 \ll (N/2)^4$. Thus, for the present case of  
spatially separated condensates the scattering rate from each
BEC is larger than the scattering rate given by $N/2$ atoms; on the other hand,
the intensity correlations are at the same time strongly reduced. These
conclusions are also valid for the case where the superpositions are of coherent
states instead of number states, as long as the overlap between the
superpositions is negligible.

Because the detection of the number state superpositions relies on atom
stimulated scattering to the BECs, the entanglement between the condensates is
not destroyed in the measurement process. However, the light scattering still
creates decoherence by phase damping explained in the previous section. This
decoherence may be reduced by balancing the detection rates of scattered light
from the BECs.

\section{Conclusions}
\label{conc}

We have shown that two BECs can be driven into macroscopic superpositions of
number and phase states by measurements of spontaneously scattered light. The
number state superpositions are entangled and spatially nonlocal
``Schr\"odinger cat" states with high occupation numbers. No stable relative
phase between the BECs is established for a finite system and in the
extreme case
the condensates approach a number state with almost all the atoms in one of the
BECs. This is an example of the strong effect of measurements on the state of
the condensates. For a finite system detections necessarily perturb the phase
and it is not irrelevant what particular phase measurement process is used;
because different measurement procedures may affect the system in a very
different way, it is not evident {\it a priori} what kind of coherence
properties, if any, are established in a detection process.

In the system considered in Ref.~{\cite{RUO97d}} two BECs are in two different
Zeeman levels and two phase coherent laser beams drive Raman transitions between
the condensates. In that case the relative phase between the two BECs is
established by measurements of spontaneously scattered photons, even though the
condensates have initially well-defined numbers of atoms. In the present paper
the large fluctuations of the number difference  between the BECs in the case of
a finite system lead to a randomly drifting relative phase. The measurement
scheme considered here, with only one laser beam and the two BECs differing in
their external quantum numbers, is closer to the experimental set-up used at MIT
in a nondestructive optical detection of a BEC \cite{AND96,AND97}. In the
phase-contrast \cite{AND97} or dark-ground \cite{AND96} imaging of the BECs the
role of the mirrors is played by a lens. Although so far all measurements of the
interference pattern of BECs have been destructive, nondestructive
measurements could possibly be performed in the near future. Only with
nondestructive imaging could one measure how the system evolves in time as a
result of the detection process.

\subsection*{Acknowledgements}
We would like to thank M. Jack and S. Tan for useful discussions. This work was
supported by the Marsden Fund of the Royal Society of New Zealand, The
University of Auckland Research Fund and The New Zealand Lottery Grants Board.
One of us (R.G.) wishes to acknowledge the hospitality of the Quantum Optics
group at the University of Auckland and support from the Deutsche
Forschungsgemeinschaft through SFB 237 'Unordnung und grosse  Fluktuationen'.

\begin{figure}
\caption{
The experimental set-up. The incoming light field is scattered from two
overlapping BECs moving with opposite velocities. The atoms scattering from one
condensate to another change the momenta of the scattered photons. The
scattered photons are collected by reflective mirrors and a 50-50 beamsplitter.
The photons are detected from the two output channels of the beamsplitter. The
photons scattered forward introduce only a constant phase shift and they are
ignored.} \label{fig:1}
\end{figure}

\begin{figure}
\caption{Stochastic simulations of the detections of spontaneously scattered
photons for 200 atoms. A typical distribution of (a) the absolute value of the
wave function $|\psi_b|$ in the condensate $b$ in the number state basis,
and (b)
the corresponding $Q$ function after approximately 5000 detections. In (a) two
distinct peaks in the number distribution correspond to entangled number state
superpositions. The peaks are centered at $N_b \simeq 30$ and at $N_b\simeq
170$
atoms. In the phase-space plotting of the $Q$ function (b) the radius in
the $xy$
plane in the polar coordinates is equal to $N_b^{1/2}$ and the polar angle
is the
relative phase between the two BECs. The four peaks correspond to the two
dominant occupation numbers and two different phase values.
 }
\label{fig:2}
\end{figure}

\begin{figure}
\caption{Another representative graph from the same run of measurements with (a)
the number state distribution and (b) the $Q$ function. The value of the
relative
phase has changed from the previous figure. The two entangled number state
superpositions and all the four phase values are clearly observed in the $Q$
representation.
 }
\label{fig:3}
\end{figure}

\begin{figure}
\caption{The same run of measurements as in previous figures. The plotting of an
extreme case in which the BECs are in an entangled number state with almost all
the atoms in one of the two condensates.
 }
\label{fig:4}
\end{figure}


\begin{references}
\bibitem{AND95} M. H. Anderson, J. R. Ensher, M. R. Matthews, C. E. Wieman, and
E. A. Cornell,  Science {\bf 269}, 198 (1995).
\bibitem{DAV95} K. B. Davis, M.-O. Mewes, M. R. Andrews, N. J. van Druten, D. S.
Durfee, D. M. Kurn, and W. Ketterle, Phys. Rev. Lett. {\bf  75}, 3969 (1995).
\bibitem{MEW96} M.-O. Mewes, M. R. Andrews, N. J. van Druten, D. M. Kurn, D. S.
Durfee, and W. Ketterle,, Phys. Rev. Lett. {\bf 77}, 416 (1996).
\bibitem{BRA97} C. C. Bradley, C. A. Sackett, and R. G. Hulet,
Phys. Rev. Lett. {\bf 78}, 985 (1997).
\bibitem{JAV96} J. Javanainen and S. M. Yoo, Phys. Rev. Lett. {\bf 76}, 161
(1996); S. M. Yoo, J. Ruostekoski, and J. Javanainen, J. Mod. Opt., in
press.
\bibitem{NAR96} M. Naraschewski, H. Wallis, A. Schenzle, J. I. Cirac, and
P. Zoller, Phys. Rev. A {\bf 54}, 2185 (1996).
\bibitem{CIR96} J. I. Cirac, C. W. Gardiner, M. Naraschewski, and P. Zoller,
Phys. Rev. A {\bf 54}, R3714 (1996).
\bibitem{JAC96} M. W. Jack, M. J. Collett, and D. F. Walls, Phys. Rev. A
{\bf 54}, R4625 (1996).
\bibitem{WON96} T. Wong, M. J. Collett, and D. F. Walls, Phys. Rev A {\bf 54},
R3718 (1996).
\bibitem{CAS97} Y. Castin and J. Dalibard, Phys. Rev. A {\bf 55}, 4330 (1997).
\bibitem{STE97} M. J. Steel and D. F. Walls, Phys. Rev. A, in press.
\bibitem{RUO97d} J. Ruostekoski and D. F. Walls, Phys. Rev. A, in press
(cond-mat/9703190).
\bibitem{JAV96b} J. Javanainen, Phys. Rev. A {\bf 54}, R4629 (1996).
\bibitem{IMA96} A. Imamo\=glu and T. A. B. Kennedy, Phys. Rev. A {\bf 55}, R849
(1997).
\bibitem{RUO97b} J. Ruostekoski and D. F. Walls, Phys. Rev. A {\bf 55}, 3625
(1997).
\bibitem{SAV97} C. M. Savage, J. Ruostekoski, and D. F. Walls,
Phys. Rev. A, in press (cond-mat/9612174).
\bibitem{AND97} M. R. Andrews, C. G. Townsend, H.-J. Miesner, D. S.
Durfee, D. M. Kurn, and W. Ketterle, Science {\bf 275}, 637 (1997).
\bibitem{GLA63} R. J. Glauber, Phys. Rev. {\bf 130}, 2529 (1963); {\bf
131}, 2766 (1963); P. L. Kelley and W. H. Kleiner, Phys. Rev. {\bf 136}, A316
(1964). 
\bibitem{CIR97} J. I. Cirac, M. Lewenstein, K. M\o lmer, and P. Zoller,
unpublished.
\bibitem{JAV95b} J. Javanainen and J. Ruostekoski, Phys. Rev. A {\bf 52}, 3033
(1995).
\bibitem{RUO97a} J. Ruostekoski and J. Javanainen, Phys. Rev. A {\bf 55}, 513
(1997); {\it ibid.}, Phys. Rev. A, in press (cond-mat/9701088).
\bibitem{DAL92} J. Dalibard, Y. Castin, and K. M\o lmer, Phys. Rev. Lett. {\bf
68}, 580 (1992).
\bibitem{GAR92} C. W. Gardiner, A. S. Parkins, and P. Zoller, Phys. Rev. A {\bf
46}, 4363 (1992).
\bibitem{CAR93} H. J. Carmichael, {\it An Open Systems Approach to Quantum
Optics}, Lecture Notes in Physics (Springer, Berlin, 1993).
\bibitem{LEG91} A. J. Leggett and F. Sols, Found. of Phys. {\bf 21}, 353 (1991).
\bibitem{WAL94} D. F. Walls and G. J. Milburn, {\it Quantum Optics} (Springer,
Berlin, 1994).
\bibitem{TIT66} U. M. Titulaer and R. J. Glauber, Phys. Rev. {\bf 145}, 1041
(1966).
\bibitem{ZUR91} W. H. Zurek, Phys. Today {\bf 44} (10), 36 (1991) and references
therein.
\bibitem{WAL85} D. F. Walls and G. J. Milburn, Phys. Rev. A {\bf 31},
2403 (1985).
\bibitem{GAR94} B. M. Garraway and P. L. Knight, Phys. Rev. A {\bf 50}, 2548
(1994).
\bibitem{AND96} M. R. Andrews, M.-O. Mewes, N. J. van Druten, D. S. Durfee,
D. M. Kurn, and W. Ketterle, Science {\bf 273}, 84 (1996).




\end{references}
\end{document}